\documentclass[aps,prb,twocolumn,groupedaddress,showpacs,psfile,floatfix]{revtex4}

\usepackage{amsbsy}
\usepackage{graphicx}
\usepackage{amsmath}

\begin{document}

\title{Green's function approach to the magnetic properties of
the kagom\'e antiferromagnet}

\author{B. H. Bernhard}
\email[]{dfi2bhb@joinville.udesc.br}
\affiliation{Dep. de F\'\i sica, Universidade do Estado de Santa Catarina,
C.P. 631, 89223-100 Joinville, SC, Brazil}
\author{B. Canals}
\email[]{canals@labs.polycnrs-gre.fr}
\author{C. Lacroix}
\email[]{lacroix@labs.polycnrs-gre.fr}
\affiliation{Laboratoire de Magn\'etisme Louis N\'eel, CNRS,
B.P. 166, 38042 Grenoble Cedex9, France}

\date{\today}

\begin{abstract}
The $S=1/2$ Heisenberg antiferromagnet is studied on the kagom\'e lattice
by using a Green's function method based on an appropriate decoupling of
the equations of motion. Thermodynamic properties as well as spin-spin
correlation functions are obtained and characterize this system as a
two-dimensional quantum spin liquid. Spin-spin correlation functions decay
exponentially with distance down to low temperature and the calculated missing
entropy at $T=0$ is found to be $0.46\ln{2}$. Within the present scheme,
the specific heat exhibits a single peak structure and a $T^2$ dependence
at low temperature.
\end{abstract}


\maketitle

\section{\label{introduction}Introduction}

Antiferromagnetic spin systems on fully frustrated
lattices show many unusual behaviors in magnetic and thermal
properties\cite{Liebmann,Ramirez}.
One of the main ingredients is that their unit cell allows for
a continuous degree of freedom and that the connectivity (corner
sharing) allows for an extensive number of these degrees of freedom
in the classical ground state.
More subtle phenomena appear when looking at quantum models on
these lattices as quantum fluctuations may induce
very unsusual ground states.
In particular, wether the quantum ground state on fully frustrated lattices
may not break the lattice symmetry nor the spin group symmetry
is still a highly debated question.
A lot of work has been done to answer this question in the
framework of the proposition of Anderson\cite{anderson}, looking first at
Resonating Valence Bond states.

One of the most studied candidates is the quantum $S$=1/2 Heisenberg
antiferromagnet on the kagom\'e lattice.
Using various methods\cite{e89,ze95,ce92,le93,lblps97,web98,ey94,mila98,si00},
it has been shown that the low temperature physics should be dominated by
short range RVB states which produce a continuum of singlet states between
the $S=0$ ground state and the first excited $S=1$ state.
A still controversial question is the presence of a very low temperature
peak in the specific heat, much below the one corresponding to the onset of short
range correlations, that could be ascribed to a high density of singlet
states in the singlet-triplet spin gap.
Associated to this low temperature peak is the missing (or not) entropy
that would characterize an ordered or a disordered ground state.

The experimental relevance of the model is extremely fragile as, in
real compounds, many other parameters may drive the physics to very
different universality classes, described through inclusion of
next nearest neighbor interactions\cite{rei91}, disorder\cite{shender93},
antisymmetric interactions\cite{elhajal01}, etc...
Nevertheless, in many cases, the Heisenberg antiferromagnet is a good
starting point as it is generally believed that many properties of
realistic systems come from deviations from the Heisenberg limit.
Examples are the layered oxide
SrCr$_{9p}$Ga$_{12-9p}$O$_{19}$\cite{rec90,ukkll94,lbar96} ($S=3/2$)
or the organic pseudo-$S=1$ compound m-MPYNN.BF$_4$\cite{wkyoya97}.

In this paper, the properties of the Heisenberg $S$=1/2 kagom\'e
antiferromagnet are addressed by using a spin Green's functions technique.
One of the advantages of the method is that it is well suited for magnetic
systems with no long range order as it uses a decoupling scheme based on
short ranged spin correlations.
This method was previously introduced by Kondo and Yamagi\cite{ky72}
in the context of the one-dimensional Heisenberg model as a theory of
spin-waves in absence of long-range order. More recently, it has been
used to address different problems\cite{sh91,yf00} and can be extended to
include magnetic phases\cite{gp02}.
Here, the formalism is used to compute thermodynamic and magnetic quantities
at all temperatures, like the internal energy, specific heat, entropy,
magnetic susceptibility and structure factor.
Correlation functions are also computed at all temperatures and for various
separation distances.
In the next section, the approximation is presented and results are discussed
in  Sec.~\ref{results}.

\begin{figure}
\includegraphics[width=5cm]{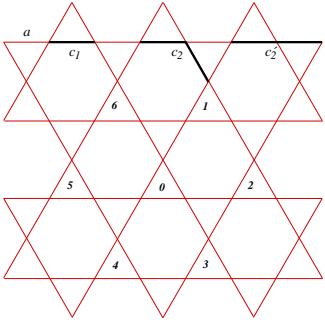}
\caption{\label{rede}The kagom\'e lattice.}
\end{figure}

\section{Model and Approximation}

The antiferromagnetic Heisenberg model is defined by the Hamiltonian
\begin{equation}
{\cal H}= \frac{1}{2}\sum _{ij} J_{ij}\ {\bf S}_i\cdot {\bf S}_j.
\end{equation}
where we assume $S=1/2$ and a nearest-neighbor exchange
\begin{equation}
J_{ij} = \begin{cases}
J (>0) & \text{if $i,j$ are nearest neighbors} \\
0 & \text{otherwise.}
\end{cases}
\end{equation}

The spin susceptibility
\begin{equation}
\chi_{ij}(\omega )\equiv \ll  S^{-}_{i}; S^{+}_{j} \gg_{\omega}
\end{equation}
is obtained by a Fourier transformed Green's function
\begin{equation}
\ll A; B\gg_\omega =
\int \ll A; B \gg _t e^{i\omega t}dt
\end{equation}
where
\begin{equation}
\ll A; B \gg _t\equiv -i\theta (t)
\left<\left[ \hat{A}(t), \hat{B}(0)\right]\right> .
\end{equation}

>From the equation of motion of the operator $\hat{A}(t)$ in the Heisenberg
representation, the Green's functions above must satisfy the equation
\begin{equation}
\omega \ll A; B\gg _\omega = \left< \left[ A, B\right] \right>
+\ll \left[ A, {\cal H}\right] ; B\gg _\omega .
\end{equation}

The spin susceptibility of the present model can be obtained by
\begin{equation}
\omega \chi_{ij}(\omega )= -2\delta _{ij} \left< S^z _i\right>
+\sum _k J_{ik} \left( \Gamma _{kij}(\omega )-\Gamma _{ikj}(\omega )\right) .
\end{equation}
The three operator Green's functions
\begin{equation}
\Gamma _{ikj}(\omega )\equiv \ll S^z _i S^- _k; S^+ _j \gg  _\omega
\end{equation}
must also obey their corresponding equations
\begin{multline}
\omega \Gamma _{ikj}(\omega )=\left( \delta _{ij} -\delta _{kj} \right)
\left< S^+ _i S^- _k\right> _\omega \\
+1/2 \sum _l J_{il} \ll \left( S^+ _i S^- _l -S^- _i S^+ _l
\right) S^- _k; S^+ _j\gg  _\omega \\
+\sum _l J_{kl} \ll S^z _i \left( S^z _l S^- _k -S^z _k S^- _l \right)
; S^+ _j\gg ,
\end{multline}
As higher-order Green's functions are generated, one obtains an infinite
hierarchy of equations which have to be decoupled in order to obtain
$\chi _{ij} (\omega )$.

In a frustrated lattice, one is constrained to the non-magnetic phase,
where $\left< S_i^z \right> =\left< S_i^\pm \right> =0$ and
$\left< S_i^+ S_j^- \right> =2 \left< S_i^z S_j^z \right> $.

Here we adopt an extension of the Kondo-Yamaji decoupling approximation
for the kagom\'e lattice, assuming (only for different sites $i, l, k$)
\begin{eqnarray}
\ll S_i^- S_l^+ S_k^- ; S_j^+ \gg _\omega & \approx & \alpha
\left< S_i^- S_l^+ \right> \chi_{kj}(\omega ) \nonumber \\
 & & +\alpha \left< S_l^+ S_k^- \right> \chi_{ij}(\omega ) ,
\label{ap1}
\\
\ll S_i^z S_l^z S_k^- ; S_j^+ \gg _\omega & \approx & \alpha
\left< S_i^z S_l^z \right> \chi_{kj}(\omega ) .
\label{ap2}
\end{eqnarray}

The static correlation functions $\left< S_i^+ S_j^- \right>$
between spins at the lattice positions ${\bf R}_i$ and ${\bf R}_j$
depend on the distance $d=|{\bf R}_i-{\bf R}_j|$, on the number
$n$ of bonds between sites $i$ and $j$, and on the lattice symmetry.
As a shorthand notation, we shall call the local correlation as
$c_0=\left<S^+ _i S^- _i\right>$, the nearest-neighbor correlation
as $c_1$ and the next-nearest-neighbor correlations as $c_2$ and $c_2'$
(the former corresponding to the shortest distance, as illustrated in
figure \ref{rede}).
We also introduce $\tilde{c}_n^{(\prime )}=\alpha c_n^{(\prime )}$.
The parameter $\alpha$ is useful to enforce the
condition $c_0=1/2$ \cite{ky72}.

These correlation functions can be evaluated from the spin susceptibility as
\begin{equation}
\left<S^+ _j S^- _i\right> = -\frac{1}{\pi}\ \int
\frac{{\rm Im}\chi _{ij} (\omega +i\eta )}
{1-e^{-\beta\omega}}\ d\omega .
\label{eq:corr}
\end{equation}
Thus, $c_1$, $c_2$, and $c_2^\prime$ and also the parameter $\alpha$
appearing in Eqs. \ref{ap1}-\ref{ap2} must be evaluated selfconsistently.

Applying approximation \ref{ap1}-\ref{ap2} to the kagom\'e lattice yields
\begin{multline}
\left[ \omega ^2-4J^2\tilde{c} \right] \chi _{ij} (\omega ) =
-8Jc_1\delta _{ij}+2c_1J_{ij} \\
+J\tilde{c}_1\sum _k J_{ik}' \chi _{kj}(\omega )
-J\left( \tilde{c}+2\tilde{c}_1 \right) \sum _k J_{ik}
\chi _{kj}(\omega ) ,
\label{eq:appr}
\end{multline}
where
\begin{equation}
\tilde{c}=1/2+\tilde{c}_1+\tilde{c}_2+\tilde{c}_2'
\end{equation}
and
\begin{equation}
J_{ij}' = \begin{cases}
J & \text{if i, j are next-nearest neighbors,} \\
0 & \text{otherwise.}
\end{cases}
\end{equation}

By working directly with the $\tilde{c}_n^{(\prime )}$, the numerical
problem is reduced to only two selfconsistent parameters, $\tilde{c}_1$
and $\tilde{c}_2+\tilde{c}_2^\prime$, from which one can obtain
$\alpha =2\tilde{c}_0$. 

With reference to the underlying triangular lattice, the approximate equation 
\ref{eq:appr} can be rewritten in matrix form
\begin{multline}
\left[ \omega ^2- 4J^2 \tilde{c} \right] \boldsymbol{\chi} _{ij}(\omega ) = 
-2c_1\left( 4J\delta _{ij}\boldsymbol{\mathit 1}-\boldsymbol{J}_{ij}\right)
\\
+J\tilde{c}_1\sum _k \boldsymbol{J}_{ik}' \boldsymbol{\chi} _{kj}(\omega )
-J \left( \tilde{c}+2\tilde{c}_1 \right) \sum _k \boldsymbol{J}_{ik}
\boldsymbol{\chi} _{kj}(\omega ) \ ,
\label{matrix}
\end{multline}
where $\boldsymbol{J}_{ij}$ and $\boldsymbol{J}'_{ij}$ are the exchange
matrices connecting neighboring triangles. As indicated in Fig. 1, each
triangle is surrounded by six neighboring triangles, all of them contributing
to both types of exchange matrices.

The elements of the dynamic susceptibility matrix are defined as
\begin{equation}
\chi ^{a b}({\bf q},\omega )= \frac{1}{N}\sum _{i,j}
\chi ^{a b}_{ij} (\omega )\ e^{-i{\bf q} \cdot
({\bf R}_i^a -{\bf R}_j^b )}
\label{dymatrix}
\end{equation}
where ${\bf R}_i^a ={\bf R}_i+{\bf T}_a$, ${\bf R}_i$ is the
lattice position of the $i$-th triangle, and the vectors ${\bf T}_a$
indicate the position of site $a$ in the triangular unit cell
($a=1,2,3$).

Applying a corresponding Fourier transformation to equation \ref{matrix} gives
\begin{equation}
\label{chi-of-q}
\boldsymbol{\chi} ({\bf q},\omega )= -2c_1
\boldsymbol{A}^{-1}({\bf q},\omega )\ \boldsymbol{B}({\bf q})
\end{equation}
where
\begin{eqnarray}
\boldsymbol{A}({\bf q},\omega ) & = &
\left( \omega ^2-4J^2\tilde{c} \right) \boldsymbol{\mathit 1}
-J\tilde{c}_1\boldsymbol{J}' ({\bf q}) \nonumber \\
 & & +J \left( \tilde{c}+2\tilde{c}_1 \right)
 \boldsymbol{J} ({\bf q}) ,
\\
\boldsymbol{B}({\bf q}) & = & 4J\boldsymbol{\mathit 1}-\boldsymbol{J}({\bf q}) ,
\end{eqnarray}
where $\boldsymbol{J} ({\bf q})$ and $\boldsymbol{J}^\prime ({\bf q})$ are
the Fourier transform of the exchange matrices
$\boldsymbol{J}_{ij}$ and $\boldsymbol{J}^\prime_{ij}$,
\begin{widetext}
\begin{equation}
\boldsymbol{J}({\bf q}) = 2J
\left( \begin{array}{ccc}
0 & {\rm cos}[(x-\sqrt{3}y)/4] & {\rm cos}[(x+\sqrt{3}y)/4] \\ \\
{\rm cos}[(x-\sqrt{3}y)/4] & 0 & {\rm cos}(x/2) \\ \\
{\rm cos}[(x+\sqrt{3}y)/4] & {\rm cos}(x/2) & 0 \\
\end{array} \right) \ ,
\end{equation}
\begin{equation}
\boldsymbol{J}^\prime ({\bf q}) = 2J
\left( \begin{array}{ccc}
2{\rm cos}(x/2){\rm cos}(\sqrt{3}y/2) & {\rm cos}[(3x+\sqrt{3}y)/4]
 & {\rm cos}[(3x-\sqrt{3}y)/4] \\ \\
{\rm cos}[(3x+\sqrt{3}y)/4] & {\rm cos}x+{\rm cos}[(x-\sqrt{3}y)/2]
 & {\rm cos}(\sqrt{3}y/2) \\ \\
{\rm cos}[(3x-\sqrt{3}y)/4] & {\rm cos}(\sqrt{3}y/2)
 & {\rm cos}x+{\rm cos}[(x+\sqrt{3}y)/2] \\
\end{array} \right) \ .
\end{equation}
\end{widetext}

\noindent in which $x=q_xa$ and $y=q_ya$.

It follows that
\begin{equation}
\boldsymbol{\chi} ({\bf q},\omega )= -2c_1
\frac{\boldsymbol{P}({\bf q},\omega ^2)}{\Delta({\bf q},\omega ^2)}
\end{equation}
where $\Delta({\bf q},\xi )$ is a third degree polynomial in $\xi$ with
${\bf q}$-dependent coefficients and the elements of
$\boldsymbol{P}({\bf q},\xi )$ are second degree polynomials
in $\xi$ with ${\bf q}$-dependent coeficients. The dynamic susceptibility
can be finally writen as
\begin{equation}
\boldsymbol{\chi} ({\bf q},\omega )=-2c_1
\sum _{l=1} ^3 \frac{\boldsymbol{R}_l({\bf q})}{\omega ^2-\Omega _l^2({\bf q})}
\end{equation}
where
\begin{equation}
\boldsymbol{R}_l({\bf q})= \frac{\boldsymbol{P}({\bf q},\Omega _l^2)}
{\Delta '({\bf q},\Omega _l^2)}
\end{equation}
and
\begin{equation}
{\Delta '({\bf q},\xi )}=\frac{\partial}{\partial\xi}\Delta ({\bf q},\xi )\ .
\end{equation}
The $\Omega _l({\bf q})$ are excitation energies, obtained from the roots
$\xi _l=\Omega _l^2({\bf q})$ of the polynomial $\Delta ({\bf q},\xi )$.
One of these roots,
$\Omega _1^2({\bf q})=6J^2(\tilde{c}+\tilde{c_1})$,
is found to be non-dispersive over the whole Brillouin zone.
Thus, the dispersion in the energy spectrum is due to the other roots
($l=2,3$), which are evaluated numerically.

By writing the correlation functions of eq. \ref{eq:corr} in matrix form
and performing the inverse Fourier transformation, we can show that
\begin{equation}
c_{ij}^{a b}=-2c_1 \sum _{l=1} ^3 \sum _{\bf q}
b_l({\bf q})\ R_l^{a b}({\bf q})\
{\rm cos}\{ {\bf q}\cdot ({\bf R}_i^a -{\bf R}_j^b )\}
\end{equation}
where
\begin{equation}
b_l({\bf q})=\frac{n(\Omega _l)-n(-\Omega _l)}{2\Omega _l},
\end{equation}
and $n(\omega )$ is the Bose distribution function.
The sum over momenta must be performed numerically.

The dynamic structure factor matrix is related to the dynamic susceptibility by
\begin{equation}
\boldsymbol{S}({\bf q},\omega )=
-\frac{2{\rm Im}\ \boldsymbol{\chi}({\bf q},\omega )}{1-e^{-\beta \omega}}.
\end{equation}
After obtaining the selfconsistent parameters, the static structure factor
can be evaluated by
\begin{equation}
\boldsymbol{S}({\bf q})=2c_1 \sum _{l=1} ^3 \sum _{\bf q}
b_l({\bf q})\ \boldsymbol{R}_l({\bf q}).
\end{equation}

\section{\label{results}Results}

\begin{figure}
\includegraphics[width=8cm]{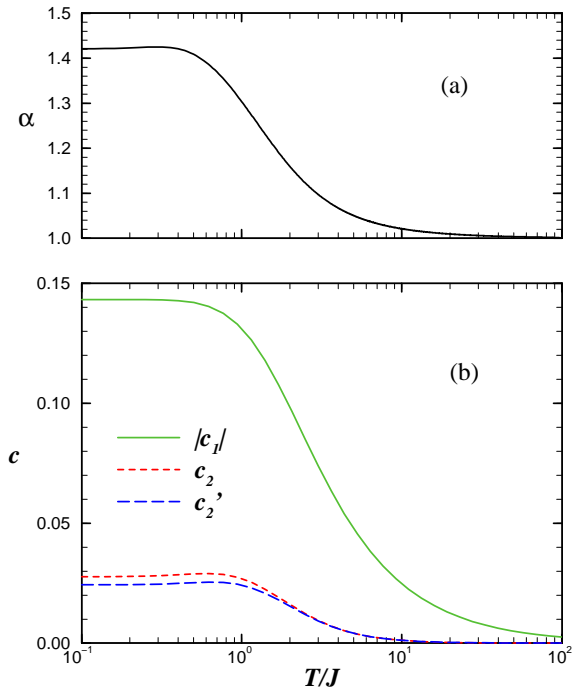}%
\caption{\label{corr}(a) The selfconsistent parameter $\alpha$ and (b) the
correlation functions $c_1$, $c_2$ and $c_2'$ as a function of temperature.}
\end{figure}

\begin{figure}
\includegraphics[width=6cm]{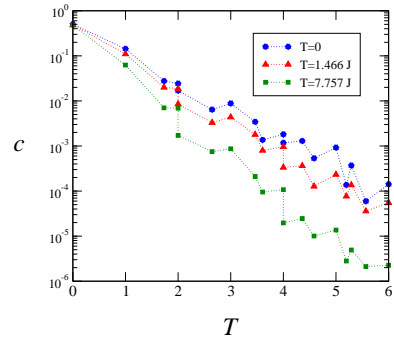}%
\caption{\label{cn}Correlation functions versus distance (in units of the lattice
parameter $a$).}
\end{figure}

We have calculated the selfconsistent values of $c_1$, $c_2$, and $c_2'$
as a function of temperature. They are plotted in fig. \ref{corr}-b. The
selfconsistent parameter $\alpha$ is shown in fig. \ref{corr}-a. From the
knowledge of these selfconsistent functions all missing correlation functions
can be evaluated. In figure \ref{cn} we show the correlation functions
$c_n^{(\prime )}$ between sites separated by at most 6 lattice bonds.
The distribution of the points is very similar to that obtained in
Ref.~\onlinecite{cl98} for the pyrochlore lattice, indicating an exponential
decay with a correlation length of the order of the lattice parameter $a$.

\begin{figure}
\includegraphics[width=6cm]{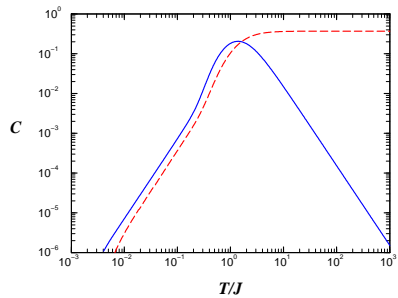}%
\caption{\label{heat}Specific heat as a function of temperature (solid line). The
dashed line gives the integrated entropy.}
\end{figure}

The internal energy is given by $E=-6NJc_1$.
The ground-state value
$E=-0.859 NJ$ is in good agreement with earlier calculations
(see Refs.~\onlinecite{lblps97,yf00,ch90}).
In figure \ref{heat} we present the specific heat
$C=dE/dT$.
This curve has a single peak and is very similar to the result of
high-temperature expansions \cite{ey94} reproducing the high-temperature
limit $C\sim T^{-2}$.
In the low-temperature limit, the time involved in the numerical computation
increases with decreasing temperature, because the convergence of the integral
in ${\bf q}$ becomes slower, and in addition one needs a higher precision in
order to perform the derivative involved in the evaluation of the specific heat.
Nevertheless, a careful analysis of the numerical results indicates
unambiguously a $T^2$ dependence of the specific heat which extends
to very low temperatures.
Therefore, there will be no second peak in the present model.
The fact that the specific heat does not vanish exponentially indicates the
presence of low energy states even though the low-temperature peak found
in Ref.~\onlinecite{si00} is missing.
Indeed, it should be emphasized that it is not sure that this
peak will persist in the thermodynamic limit\cite{sin01} which
supports the results obtained in the present work.
The change in entropy $\int _0 ^T\frac{C(T^\prime)}{T^\prime}dT^\prime$
is also shown in Fig. \ref{heat}.
The total change in entropy is $0.54\ln{2}$,
corresponding to a ground-state entropy $0.46\ln{2}$, of the same order
as found in Ref.~\onlinecite{ey94}.

\begin{figure}
\includegraphics[width=6cm]{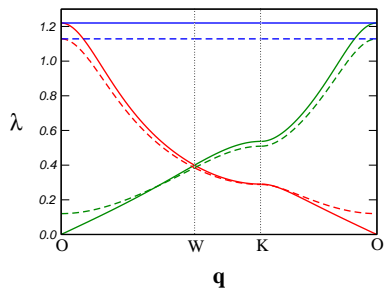}%
\caption{\label{lambda}The eigenvalues $\lambda _n({\bf q})$ of the structure
factor along some symmetry directions in ${\bf q}$-space at $T=0$ (solid lines)
and for an intermediate value of $T$ (dashed lines).}
\end{figure}

The eigenvalues $\lambda _n({\bf q})$ of the structure factor are shown in
Fig. \ref{lambda}. The result is remarkably similar to what is found from
the high-temperature expansions of Ref. \cite{ey94}. Whithin the present
approach one can easily follow the effect of temperature on the structure
factor for all wavevectors even at very low $T$. We observe that its highest
eigenvalue is degenerate over the whole Brillouin zone.
For the pyroclore, the structure factor is a $4\times 4$ matrix, which has
one additional eigenvalue. In a perturbative expansion \cite{cl98} the two
highest eigenvalues are found to lie very close to each other, one of
them being completely degenerate while the other one has a weakly lifted
degeneracy, which is crucial to reproduce the main features of neutron
scattering experiments. As a further work, it would be interesting to
extend the present approach to the pyrochlore antiferromagnet.

\begin{acknowledgments}
This work has been partially supported by Brazilian
agencies CNPq (Conselho Nacional de Desenvolvimento Cient\'\i fico e
Tecnol\'ogico) and CAPES (Coordena\c c\~ao de Aperfei\c coamento de
Pessoal de N\'\i vel Superior) through the French-Brazilian cooperation
agreement CAPES-COFECUB.
\end{acknowledgments}


\begin{thebibliography}{10}

\bibitem{Liebmann} R.~Liebmann:
{\it Statistical Mechanics of Periodic Frustrated Ising Systems},
(Springer, Berlin, 1986).

\bibitem{Ramirez}
A.~P. Ramirez, Ann. Rev. Mater. Sci {\bf 24},  453  (1994), P. Schiffer and
A.~P. Ramirez, Comments Condens. Matter Phys. {\bf 18}, 21 (1996) and references
  therein.

\bibitem{anderson}
P. W. Anderson, B. Halperin and C. M. Varma,
Phisos. Mag. {\bf 25}, 1 (1972).

\bibitem{e89}
V. Elser, Phys. Rev. Lett. {\bf 62},  2405  (1989).

\bibitem{ze95}
C. Zeng and V. Elser, Phys. Rev. B {\bf 51},  8318  (1995).

\bibitem{ce92}
J. Chalker and J. Eastmond, Phys. Rev. B {\bf 46},  14201  (1992).

\bibitem{le93}
P. Leung and V. Elser, Phys. Rev. B {\bf 47},  5459  (1993).

\bibitem{lblps97}
P. Lecheminant {\it et~al.}, Phys. Rev. B {\bf 56},  2521  (1997).

\bibitem{web98}
C. Waldtmann {\it et~al.}, Eur. Phys. J. B {\bf 2},  501  (1998).

\bibitem{ey94}
N. Elstner and A.~P. Young, Phys. Rev. B {\bf 50},  6871  (1994).

\bibitem{mila98}
F. Mila, Phys. Rev. Lett. {\bf 81}, 2356 (1998).

\bibitem{si00} P. Sindzingre, G. Misguich, C. Lhuillier, B. Bernu, L. Pierre,
Ch. Waldtmann, and H.-U. Everts, Phys. Rev. Lett. {\bf 84}, 2953 (2000).

\bibitem{rei91}
J. N. Reimers, A. J. Berlinsky and A.-C. Shi.,
Phys. Rev. B {\bf 43} (1991) 865.

\bibitem{shender93}
E. F. Shender, V. B. Cherepanov, P. C. W. Holdsworth, and A. J. Berlinsky,
Phys. Rev. Lett. {\bf 70}, 3812-3815 (1993).

\bibitem{elhajal01}
M. Elhajal, B. Canals, C. Lacroix, {\it to be published}.

\bibitem{rec90}
A. Ramirez {\it et~al.},   Phys. Rev. Lett. {\bf 64},  2070 (1992).

\bibitem{ukkll94}
Y. Uemura {\it et~al.}, Phys. Rev. Lett. {\bf 73},  3306  (1994).

\bibitem{lbar96}
S.-H. Lee {\it et~al.}, Europhys. Lett {\bf 35},  127  (1996).

\bibitem{wkyoya97}
N. Wada {\it et~al.}, J. Phys. Soc. Jpn. {\bf 66},  961  (1997).

\bibitem{ky72}
J. Kondo and K. Yamaji, Prog. Theor. Phys. {\bf 47}, 807 (1972).

\bibitem{sh91}
H. Shimahara, S. Takada, J. Phys. Soc. Jap. {\bf 60}, 2394 (1991).

\bibitem{yf00} W. Yu and S. Feng, Eur. Phys. J. B {\bf 13}, 265 (2000).

\bibitem{gp02} M. E. Gouv\^ea and A. S. T. Pires, {\it to be published}

\bibitem{cl98}
B. Canals and C. Lacroix, Phys. Rev. Lett. {\bf 80}, 2933 (1998).

\bibitem{ch90}
Chen Zeng and Veit Elser, Phys. Rev. B {\bf 42}, 8436 (1990).

\bibitem{sin01}
P. Sindzingre, {\it private communication}.

\end{thebibliography}
\end{document}